\documentclass[12pt]{article}   
\usepackage{psfig,times,fancyheadings}  

\begin{document}
\thispagestyle{empty} 

\baselineskip=20pt  
 \lhead[\fancyplain{}{\sl }]{\fancyplain{}{\sl }}
 \rhead[\fancyplain{}{\sl }]{\fancyplain{}{\sl }}

 \renewcommand{\topfraction}{.99}      
 \renewcommand{\bottomfraction}{.99} 
 \renewcommand{\textfraction}{.0}


\newcommand{\nc}{\newcommand}

\nc{\qI}[1]{\section{{#1}}}
\nc{\qA}[1]{\subsection{{#1}}}
\nc{\qun}[1]{\subsubsection{{#1}}}
\nc{\qa}[1]{\paragraph{{#1}}}

\def\qbu{\hfill \par \hskip 6mm $ \bullet $ \hskip 2mm}
\def\qee#1{\hfill \par \hskip 6mm #1 \hskip 2 mm}

\nc{\qfoot}[1]{\footnote{{#1}}}
\def\qL{\hfill \break}
\def\qpar{\vskip 2mm plus 0.2mm minus 0.2mm}
\def\qtvi{\vrule height 2pt depth 5pt width 0pt}
\def\qth{\vrule height 12pt depth 0pt width 0pt}
\def\qtb{\vrule height 0pt depth 5pt width 0pt}
\def\tvi{\vrule height 12pt depth 5pt width 0pt}

\def\qparr{ \vskip 1.0mm plus 0.2mm minus 0.2mm \hangindent=10mm
\hangafter=1}

\def\qdec#1{\par {\leftskip=2cm {#1} \par}}

\def\qdpt{\partial_t}
\def\qdpx{\partial_x}
\def\qddpt{\partial^{2}_{t^2}}
\def\qddpx{\partial^{2}_{x^2}}
\def\qn#1{\eqno \hbox{(#1)}}
\def\qds{\displaystyle}
\def\qw{\widetilde}
\def\qmax{\mathop{\rm Max}}   
\def\qmin{\mathop{\rm Min}}   


\def\qci#1{\parindent=0mm \par \small \parshape=1 1cm 15cm  #1 \par
               \normalsize}

\null
\vskip 0.1 cm

\centerline{\bf \Large Consensus formation:}
\vskip 2mm
\centerline{\bf \Large The case of using cell phones while driving}                                      

\vskip 7mm
\centerline{\bf Bertrand M. Roehner $ ^1 $ }
\vskip 3mm
         
\centerline{\bf Institute for Theoretical and High Energy Physics}
\centerline{\bf University Paris 7 }

\vskip 1cm

{\bf Abstract}\quad Several models (including the widely used Sznajd model)
have been proposed in order to describe the social phenomenon of consensus
formation. The objective of the present paper is to supplement the simulations
based on these models with a ``real world simulation''; it considers a situation
which can be considered as an ideal laboratory for analyzing consensus formation,
namely the authorization or prohibition of using cell phones while
driving. This is a convenient laboratory for several reasons 
(i) The issue was raised
in similar terms in all industrialized countries, a circumstance that
facilitates comparative
analysis by providing a {\it set of observations} (as opposed to
single observations generated, for instance, in election contests). 
(ii) This is a situation where we happen to know the rule
a ``rational'' agent should follow. 
(iii) Because the issue
is a matter of life and death, the phenomenon can be considered as
fairly robust with respect to various, endogenous or exogenous,
sources of noise. 
(iv) The relevant data are 
available on the Internet and in newspaper data bases. 
\qL
Our observations strongly suggest that there is a missing variable
in most consensus formation simulations. In its conclusion, the paper calls 
for a large scale effort for identifying, documenting and analyzing other 
real-world cases of consensus formation.

\vskip 0.3cm

\centerline{October 11, 2004}

\vskip 2mm

\vskip 0.3cm
Key-words: consensus formation, Sznajd model, cell phone,
traffic fatalities.
\vskip 0.5cm 

1: ROEHNER@LPTHE.JUSSIEU.FR
\qL
\phantom{1: }Postal address where correspondence should be sent:
\qL
\phantom{1: }B. Roehner, LPTHE, University Paris 7, 2 place Jussieu, 
F-75005 Paris, France.
\qL
\phantom{1: }E-mail: roehner@lpthe.jussieu.fr
\qL
\phantom{1: }FAX: 33 1 44 27 79 90

\vfill \eject

\qI{Introduction}

In recent years the questions of information diffusion and consensus
formation were studied by many authors. Among the landmark papers
let us mention Kacperski et al. (1996), Sznajd-Weron et al. (2000), 
Stauffer et al. (2000), Stauffer (2001, 2003), Ball (2003),
Behera et al. (2003). The last paper 
provides a clear comprehensive
overall view of the applications of cellular automata
in voting and consensus formation problems.
This paper differs from the rich sample of simulations already
proposed in the sense that it is what can be called a ``real world simulation''.
Naturally, the data in table 2 (below) don't have
the precision of simulated data, but the trend that they illustrate is
sufficiently different from what simulations based on the
Sznajd model predict to foster further reflection. In a nutshell, the paper's 
main message is that socio-political macro-factors
may, to a considerable degree, distort the standard pattern
of consensus formation, either by restraining or by amplifying
the diffusion of information. 
We use the expression 
``macro-factors'' because usually they concern a whole community,
region or country. 
Although this paper focuses on
a specific case-study, namely the question of using a mobile phone
while driving, there is good evidence that the existence of macro-factors 
is the rule rather 
than the exception (see the concluding section). 
Naturally, it would not be difficult to insert macro constraints
into a one- or two-dimensional lattice simulation (they could be
seen as additional external fields). 
The problem
is rather that through
the introduction of an extra set of unknown parameters
the comparison between the model and the real world would lose
much of its relevance, at least
unless these parameters can be estimated
{\it a priori}. In short, this problem calls for developing a reliable 
methodology for estimating macro constraints. 
This may not be an easy task, however.
\qpar

The paper proceeds as follows. The second section provides
experimental and observational evidence about the risk of 
using a cell phone while driving. The third section examines how
this information spread among industrialized countries. 
It reveals great differences: in some countries the  process took
one or two years, in others it took over ten years. In the
latter cases, obviously the propagation of information was 
hindered and restrained by some powerful constraints. 
These constraints were able to block transmission 
through formal mass media channels as well as through informal
channels (rumors, hearsay, gossip).
In the conclusion,
we briefly address the question of the origin and strength of
those macro-constraints. 
\qpar

Before getting started it should be emphasized that what makes this
comparative study possible is the fact that the phenomenon
that we consider takes comparable forms in different countries. Cell phones,
cars, the reactions of drivers to disturbances, 
roads, traffic rules, or mass-media channels
are fairly similar in all industrialized countries.
It is this uniformity and 
homogeneity, that makes cross-national comparisons meaningful%
\qfoot{Naturally, there are also  factors which are {\it not} identical.
This, in itself, is not an obstacle to 
cross-national comparisons however. 
A physical parallel may help to explain why. 
When the Foucault pendulum experiment is performed
for instance in Stockholm and in Rome, the pendulums which are
used are certainly not {\it identical}; their lengths, masses, the nature of
their suspensions may not be the same, but this does not 
prevent the comparison of the observations to be significant 
because it is known (from former pendulum experiments)
that these factors do not critically
affect the phenomenon under consideration;
whatever influence they may have, it will be at least one order of 
magnitude smaller than the error bars of the main observation and
can therefore be ignored.}%
.

\qI{The evidence}

In the {\it Wall Street Journal} of July 19, 2004 we read that wireless 
companies ``point out to research that could bolster the notion
that use by drivers is not a big problem''. Similarly, {\it RER Wireless
News} of May 2002 observes that ``we are still in the information
collecting stage''. The last statement may appear surprising if
one recalls that cell phones have been in use at least since 1995
and that radio telephones have been used in cars for much 
longer. As a matter of fact, as shown in table 1, the question
has been actively investigated by academic
researchers and safety agencies, leading to fairly clear and
irrefutable evidence. 
\qpar

One of the most unquestionable observations was made in Japan.
The ban on using hand-held cell phones while driving was imposed
in November 1999. In the 6 months before enforcement there
were 1,473 traffic accidents connected 
with drivers using mobile phones, whereas
in the 6 months after the ban there were only 580 which represents a 
61\% drop (ROSPA p. 18).
\qpar

The other conclusions which emerge from table 1 are the following.
\qbu The risk of accident is multiplied by a factor of the order of
4 to 5 when a cell phone is used while driving. 
\qbu Hands-free phones offer no benefit. A study by researchers
at the university of Toronto and published in the New England 
Journal of Medicine (February 1997)  
authorized a comparison between hand-held
phones and hands-free phones. The risk was found to be
4 times higher for the first device and 6 times higher for the second.
This result could seem fairly counter-intuitive but becomes more
understandable when one realizes that people with hands-free
phones tend to make longer calls than people with 
hand-held phones and all studies have shown that it not the
fact of holding the wheel with only one hand which is dangerous
but rather the fact that the driver's attention is captured by the
conversation. In short, hands-free phones do not provide any
benefit in terms of safety. 
\qbu A study published in early 2004 by the Harvard Center of
Risk Analysis estimated that drivers talking on cell phones are
responsible for about 6\% of all US auto accidents each year.
This represents 2,600 people killed (i.e. 10 fatalities
per million of population) 
and 330,000 injured (Sundeen 2004, p. 3).
\qpar



\begin{table}[htb]

 \small 

\centerline{\bf Table 1\quad Risk involved in using cell phones while
driving}

\vskip 3mm
\hrule
\vskip 0.5mm
\hrule
\vskip 2mm

$$ \matrix{
\tvi 
 &\hbox{Year}  & \hbox{Country} \hfill& \hbox{Main conclusions} \hfill \cr
\noalign{\hrule}
\qth 
1 & 1995 & \hbox{France} \hfill & \hbox{Reaction time 60\% slower} \hfill \cr 
 &  & \hbox{} \hfill & \hbox{} \hfill \cr 
 2 &  1997 & \hbox{Canada} \hfill & \hbox{Risk of accident multiplied 
by 4 to 5; hands-free phones offer no benefit} \hfill \cr 
 &  & \hbox{} \hfill & \hbox{} \hfill \cr 
 3 & 1999 & \hbox{Japan} \hfill & \bullet \ \hbox{In the 6 months before the ban
there were 1,473 cell phone related} \hfill \cr 
 &  & \hbox{} \hfill & \hbox{accidents; in the 6 months after the ban there were
580} \hfill \cr 
 &  & \hbox{} \hfill & \bullet \ \hbox{In the 12 months before the ban
there were 2,830 cell phone related} \hfill \cr 
 &  & \hbox{} \hfill & \hbox{accidents; in the 12 months after the ban there were
1,391} \hfill \cr 
 &  & \hbox{} \hfill & \hbox{} \hfill \cr 
 4 &  2000 & \hbox{UK} \hfill & \hbox{Fourfold risk during (and up to 5 mn after)
cell phone calls} \hfill \cr 
 &  & \hbox{} \hfill & \hbox{} \hfill \cr 
 5 & 2001 & \hbox{US (Utah)} \hfill & \hbox{Phone conversations create distractions
levels much higher than other} \hfill \cr 
 &  & \hbox{} \hfill & \hbox{activities such as radio, talking with
passengers, etc.} \hfill \cr 
 &  & \hbox{} \hfill & \hbox{} \hfill \cr 
 6 &  2004 & \hbox{US (Harvard)} \hfill & \hbox{Cell phone related accidents 
kill each year 2,600 people in the} \hfill \cr 
 &  & \hbox{} \hfill & \hbox{United States (330,000 injured)} \hfill \cr 
 &  & \hbox{} \hfill & \hbox{} \hfill \cr 
\qtb
 7 &  2004 & \hbox{Sweden} \hfill & \hbox{No significant difference between
hand-held and hands-free phones } \hfill \cr 
\noalign{\hrule}
} $$

\vskip 1.5mm
Notes: The country(third column) is where the study was carried out.
Although
in a general way the data published by US statistical
agencies are at top level in terms of coverage and availability, for
this particular problem American publications  lagged behind those
of other industrialized countries. It should be noted that Japan was one of
the few countries where police reports gave indications about cell phone
use in accidents; in most other countries this information is not 
recorded.
\qL
Sources: 1: Le Monde (29 Dec. 1995); 2: New England Journal of Medicine
(February 1997) cited in Le Figaro (13 Feb. 1997); 3: Edmonton Sun (Alberta,
Canada, 27 Feb. 2004); 4: The Independent (6 May 2000), Daily Telegraph
(4 Oct. 2003); 5: Daily Mail (30 Jan. 2003); 6: Edmonton Sun (Alberta,
Canada, 27 Feb. 2004); http://www.ncsl.org;
7: International Herald Tribune (12 Apr. 2004).
\vskip 2mm

\hrule
\vskip 0.5mm
\hrule

\normalsize

\end{table}


In order to realize that the previous data are 
consistent with one another and to make them fit with intuition,
one should add a few explanations.
The main question is how the fourfold ratio which was observed
in the Canadian study can be consistent with the {\it twofold} decrease in 
the number of accidents that what recorded in Japan?
\qee{1)} First, one should note that only hand-held phones were 
prohibited in Japan; if all kinds of phones had been prohibited
a sharper decrease would likely have been observed. 
\qee{2)} Secondly, if (as claimed by the Toronto study)
hands-free phones are not safer than hand-held phones one
may wonder how the decrease observed in Japan should be
accounted for. The explanation is certainly that back in 1999
only few drivers had got hands-free sets which means that the vast
majority had to stop using their phones altogether until having been
able to buy hands-free sets. If this interpretation is correct one would
expect the number of accidents to increase in the course of time
as more and more drivers are able to use hands-free phones. 
This is indeed confirmed by the fact that in the second of the 6-month
interval after the interdiction, 
the reduction in the number of accidents was
smaller than over the first 6-month interval: it fell from 61\% to 43\%. 
(Edmonton Sun, Alberta, Canada February 27, 2004).
In other words, the benefits of the ban in terms of safety 
eroded in the course of time.
\qee{3)} From a practical and intuitive point of view one may wonder
why conversations on the phone are more distracting than
conversations with passengers. There are two reasons for that.
(i) Passengers spontaneously stop talking when the driver faces a 
difficult situation, for instance during a tricky overtaking. 
(ii) There is evidence
that phone conversations have a high emotional content. After all, as cell phone
call are fairly expensive, people usually make a call in order to say something
important. An observation which support this assertion is the fact
that the risk remains significantly 
higher over a time interval of at least 15 minutes after the end of the
conversation. In short, cell phone conversations indeed seem to have a higher
emotional content than conversations with passengers.
\qpar

In the next section we analyze to what extent the 
information summarized in table 1 was disseminated in various countries.

\qI{Information dissemination and consensus formation}

The dissemination of information by medias, road safety agencies,
automobile clubs, etc. can be studied in two different ways.
(i) By trying to assess the content of newspaper articles, agency reports, etc.
(ii) By examining the legislation which was enacted to cope with
the problem. The first methodology is difficult to implement, not only because
of the great number of publications, but also because it is not easy
to assess their content in an objective and quantitative way. In the second
approach, one considers only the outcome in terms of new legislation;
this is a kind of back box approach in which the input is the evidence
presented in table 1 and the output the legislation summarized in
table 2. Although much of this paper relies on this black box approach,
in the appendix we also summarize some of the articles to get a more
concrete feeling. 
\qpar

As one knows, the Sznajd model always leads to complete consensus.
On account of the evidence presented in table 1, the only ``rational''
consensus would be to forbid the use of cell phones in cars, whether
hand-hold or hands-free. This would safe at least 5,000 lives annually
in North America and Europe and a much larger number in coming years
as cell phones and cars become more common in countries with large
populations such as China, India and Indonesia. However, as shown
in table 2, not a single country has banned hands-free phones. This 
suggest that, although available, the information pertaining to the risk
of hand-free phones has been little circulated. In contrast, the information
about the risk of hand-held phones was disseminated in many countries
but with great differences in the delay required for consensus 
formation. 
\qpar



\begin{table}[htb]

 \small 

\centerline{\bf Table 2\quad Ban on cell phone use while driving: cross-national
comparison}

\vskip 3mm
\hrule
\vskip 0.5mm
\hrule
\vskip 2mm

$$ \matrix{
\tvi 
 &\hbox{Country}  \hfill & \hbox{Year}  & \hbox{Month} \hfill \cr
\noalign{\hrule}
\qth 
1 &\hbox{Austria}  \hfill & 2002 & \hbox{} \hfill \cr
2 &\hbox{Canada}  \hfill & \hbox{No ban} & \hbox{} \hfill \cr
 3 &\hbox{Denmark}  \hfill &  1998 & \hbox{July} \hfill \cr
 4 &\hbox{Finland}  \hfill & 2003 & \hbox{January} \hfill \cr
 5 &\hbox{France}  \hfill &  2003 & \hbox{April} \hfill \cr
 6 &\hbox{Germany}  \hfill &  2001 & \hbox{February} \hfill \cr
 7 &\hbox{Italy}  \hfill &  2003 & \hbox{July} \hfill \cr
 8 &\hbox{Japan}  \hfill & 1999 & \hbox{November} \hfill \cr
 9 &\hbox{South Korea}  \hfill &  2001 & \hbox{July} \hfill \cr
 10 &\hbox{Spain}  \hfill & 2002 & \hbox{} \hfill \cr
 11 &\hbox{Sweden}  \hfill & \hbox{No ban} & \hbox{} \hfill \cr
 12 &\hbox{UK}  \hfill & 2003 & \hbox{December} \hfill \cr
\qtb
 13 &\hbox{US}  \hfill & \hbox{No ban} & \hbox{} \hfill \cr
\noalign{\hrule}
} $$

\vskip 1.5mm
Notes: The bans concern hand-held phones only; so far no country has banned
hands-free phones in spite of reliable evidence showing
that using them is no less dangerous. ``No ban'' means that (as of
July 2004) there has been no nationwide ban. 
In New York State,
overriding Mayor Michael Blomberg's veto, the City Council has passed a ban
which became effective in November 2001. However, ticketed drivers can
escape the \$ 100 penalty if they can prove that they have bought a hands-free
set since they were stopped. In July 2004, New Jersey became the second state
with a ban on hand-held phones. In this case, however, police may charge
violators only after stopping them for another infraction. In European
states, the penalty ranges from 30 euros in  Germany to 60 euros in 
Denmark. 
\qL
Sources: Guardian (23 March 2002); Agence France Presse (1 December 2003);
Miami Herald (8 March 2004); http://www.cellular\_news.com/car\_bans; 
http://www.nj.com/printer.
\vskip 2mm

\hrule
\vskip 0.5mm
\hrule

\normalsize

\end{table}


{\bf A missing variable} \quad
In the Sznajd model each person holds one of several opinions
and attempts to induce the same opinion  in its neighbors on 
a checkerboard lattice. Various rules describe this persuasion 
process. For example, if two people adopt the same position,
all their neighbors follow suit. In the course of time, the distribution
of opinions fluctuates across the board until a consensus is reached
with all sites sharing the same view. 
For most of the countries listed in table 2
there is an initial kernel of people holding opinion X (e.g. that 
using cell phone while on the wheel is risky) composed (at the very least)
of the employees of safety agencies whose job it is to make
risk assessments. In each country, the technical means by which
information can be disseminated are very much the same, and yet,
we have these huge differences in the delays required for passing 
new legislation. It could be argued that legislation is more difficult
to pass in countries which are more centralized, such as Britain, France
or Japan, because it takes longer for the information to reach 
the top of the ``pyramid''. However, we observe the opposite:
in the United States where such legislation can be passed at state
level, the process took longer (even at the level of individual states)
than in countries where
the decision had to be taken at national level. In short, in their present
form, consensus formation models can hardly account for what is
observed. There is a missing variable. 

\qI{Conclusion}

The observation that the diffusion of news is affected by socio-economic, political
or ideological factors comes hardly as a surprise. In the present case however,
the information is not merely a piece of
news, it is a matter of life and death. One might
expect that in such a case the information would be transmitted as quickly as
possible. If macro-factors play a crucial role even in such a case, one must
wonder in which kind of situations ``pure'' consensus formation models (such as
the Sznajd model) may apply. In order to bridge the gap between theory
and observation, great efforts should be devoted to identifying, documenting
and analyzing such situations. 
\qpar

The present example can also shed some light on the phenomenon of
speculative bubbles. In 1999-2000 when the average price earnings ratio of
the NASDAQ market reached the level of 200 (i.e. 10 times higher than 
its long-term average) it should have been obvious to everybody
that the market was deeply out of balance. Yet, many medias claimed that
the situation was sustainable. ``We are making history'' was the motto at
that time. As in the cell phone case, the ``rational'' opinion was
overridden by a flood of opinions generated by a collective phenomenon
of self-delusion.
\qL
Currently (mid-2004) we have a similar situation in real estate markets.
After the more than 100\% price increase that took place in many large cities
(e.g. Boston, London, San Francisco, Sydney) it should be obvious that over
the next five years (real) prices will fall by about the same percentage. Yet,
real estate experts and medias are almost unanimous in claiming 
that the price decrease will not exceed 15\%.
In short, 
these are examples of what can
be called collective self-suggestions%
\qfoot{Perhaps the diffusion of such fallacies could provide
a suitable field of application for ``pure'' consensus formation models.}%
.
\qpar

One feature shared by all these cases is that
huge amounts of money are at stake. For instance, according to an estimate
given by the Wall Street Journal (19 July 2004) for the
United States 40\% of the traffic on
cellular phones is due to drivers; with a total revenue of US wireless carriers
of the order of \$ 90 billions in 2003, the share of drivers is \$ 37 billions.
\qpar

\appendix

 \qI{Appendix: Qualitative evidence about smoke screens}

How do medias manage to duck the real issues?
Let us for a moment return to the {\it Wall Street Journal} article that we mentioned
in section 2. This was a full-page paper and 
one can therefore wonder how the authors were able
to give a substantial account without ever mentioning the
simple and plain facts reported in table 1.
The question is of some interest because the arguments and
tricks used as red herrings and smoke screens are not specific to this paper, but
are very much the same in all the publications that try to sidestep the 
problem.
First, the bias of the papers is reflected in the terms used by the authors:
the word
``safety'' appears 15 times, whereas the word ``accident'' appears only once
and the words ``fatalities'' or ``death'' do not appear at all. The real challenge
of such a paper is to give the impression that the issue is taken seriously,
yet without mentioning hard facts. 
One trick is to give the impression that it is a
complex problem for which no clear conclusion can be reached and to embed it
into a variety of other issues (``a long list of other potential distractions, 
such as unruly children and talkative passengers''). Another artifice is to 
rely on studies which seem to support the view that there is no problem:
``In the summer of 2003, a study by University of North Carolina 
 researchers appeared to suggest that cell-phone use by drivers wasn't a 
big problem. 
The AAA [American Automobile Association] Foundation for Traffic Safety which
funded the research, declared cellphones ranked next to last on a list
of common distractions for drivers. The cellular-industry association issued
a news release, and 
newspapers nationwide reported the study
as evidence that cellphones are a minor distraction.''
How can one understand that a study performed by academic researchers
lead to results which are so blatantly at variance with those summarized
in table 1. The answer consists in two points (i) The complete release of the
North Carolina study contained reservations stating that
the findings have a number of serious limitations.
Yet, these limitations were not reported in the website posting of the
AAA Foundation for Traffic Safety, nor were they mentioned in the
newspaper reports. (ii) One should recall that the AAA is hardly an impartial 
player on this matter
since its local sections sell cellphone services (an information
which is given in the {\it Wall Street Journal}  article albeit at the very end).
Finally, the final argument consists in saying 
that it would be useless to pass
new laws limiting cell-phone use because it would be impossible to
enforce them anyway
due to budget limitations (see the Miami Herald March 8, 2004, that 
argument is cited in many other publications as well). 
\qpar

A testimony of the success of such disinformation campaigns is the fact that,
according the the National Conference
of State Legislation, 31 US states tried to introduce cellphone driving laws
in 2003 but none succeeded. With none of the European countries having
banned the use of hands-free phones, the situation is not very
different in Europe.

\vskip 10mm

\centerline{\large \bf References}
\vskip 5mm

\qparr
Ball (P.) 2002: Third beats second in vote. 
Nature 27 February 2002.

\qparr
Behera (L.), Schweitzer (F.) 2003: On spatial consensus formation:
is the Sznajd model different from a voter model?
International Journal of Modern Physics C 14, 10, 1331-1354.

\qparr
Kacperski (K.), Holyst (J.A.) 1996: Phase transition and hysteresis
in a cellular automata-based model of opinion formation.
Journal of Statistical Physics 84, 169-189. 

\qparr
ROSPA (The Royal Society for the Prevention of Accidents) 2002:
The risk of using a mobile phone while driving.
Transport Local Government and Regions. 

\qparr
Sznajd-Weron (K.), Sznajd (J.) 2000: Opinion evolution in 
closed community.
International Journal of Modern Physics C, 11, 6, 1157-1165.

\qparr
Stauffer (D.), Sousa (A.O.), de Oliviera (S.M.) 2000: Generalization to
square lattice of Sznajd sociophysics model. 
International Journal of Modern Physics C 11, 6, 1239-1245.

\qparr
Stauffer (D.) 2001: Monte Carlo simulations of Sznajd models.
Journal of Artificial Societies and Social Simulation 5, 1.

\qparr
Stauffer (D.) 2003: How to convince others? Monte Carlo simulations 
of the Sznajd model. In: The Monte Carlo method in the physical
sciences, J. Gubernatis ed. AIP Conference Proceedings.

\qparr
Sundeen (M.) 2004: Cell phones and highway safety: 2003 state legislative
update. National Conference of State Legislatures (January 23, 2004). 
This document can be found on the following website:
http://www.ncsl.org/programs/transportation/cellphoneupdate1203.htm.

\end{document}